\documentclass[layout = twocolumn,journal = jacsat]{achemso}
\setkeys{acs}{usetitle = true}

\usepackage{graphicx}
\usepackage[usenames]{color}
\usepackage{amssymb}
\usepackage{amsmath}
\usepackage{amsfonts}
\usepackage{mathrsfs}

\renewcommand{\epsilon}{\varepsilon}
\newcommand{\figurewidth}{0.45\textwidth}
\newcommand{\narrowfigurewidth}{0.45\textwidth}

\author{Wancheng Yu}

\affiliation{CAS Key Laboratory of Soft Matter Chemistry, Department of Polymer
Science and Engineering, University of Science and Technology of China, Hefei, Anhui
Province 230026, P. R. China}

\author{Kaifu Luo}
\email{kluo@ustc.edu.cn}
\affiliation{CAS Key Laboratory of Soft Matter Chemistry, Department of Polymer
Science and Engineering, University of Science and Technology of China, Hefei, Anhui
Province 230026, P. R. China}

\title [An \textsf{achemso} demo] {Chaperone-assisted translocation of a polymer through a nanopore}

\begin{document}


\begin{abstract}

Using Langevin dynamics simulations, we investigate the dynamics of chaperone-assisted translocation of a flexible
polymer through a nanopore. We find that increasing the binding energy $\epsilon$ between the chaperone and the chain and the chaperone concentration $N_c$ can greatly improve the translocation probability. Particularly, with increasing the chaperone concentration a maximum translocation probability is observed for weak binding. For a fixed
chaperone concentration, the histogram of translocation time $\tau$ has a transition from long-tailed distribution to Gaussian distribution with increasing $\epsilon$. $\tau$ rapidly decreases and then almost saturates with increasing binding energy for short chain, however, it has a minimum for longer chains at lower chaperone concentration. We also show that $\tau$ has a minimum as a function of the chaperone concentration. For different $\epsilon$, a nonuniversal dependence of $\tau$ on the chain length $N$ is also observed. These results can be interpreted by characteristic entropic effects for flexible polymers induced by either crowding effect from high chaperone concentration or the intersegmental binding for the high binding energy.


\end{abstract}


\maketitle

\section{Introduction}

The transport of biopolymers through a nanopore embedded in a membrane has attracted wide attention because it closely
connects with polymer physics and is also related to many crucial processes in biology, examples including the passage
of DNA and RNA through nuclear pores, the translocation of proteins through the endoplasmic reticulum, as well as the
viral injection of DNA into a host \cite{Alberts}. In addition, the translocation processes have been suggested to have potentially revolutionary technological applications, such as rapid DNA or RNA sequencing \cite{Kasianowicz,Meller,Branton}, gene therapy \cite{Hanss} and controlled drug delivery \cite{Holowka}.

Polymer translocation through a nanopore faces a large entropic barrier due to the loss of a great number of available
conformations, thus driving forces are introduced. Two important driving forces for translocation, both in experimental setups and $in$ $vivo$, are provided by an electric field across the membrane and binding proteins (so-called chaperones).
Translocation driven by an electric field has recently been investigated extensively \cite{Sung,Muthukumar,Han,Kantor,Luo,Luo4,Luo1}.
In this study, we focus on the latter driving mechanism, which is responsible for the translocation of
proteins \cite{Matlack,Brunner} as well as the DNA translocation through membranes \cite{Salman,Farkas}.
Up to now, there are also several theoretical and numerical studies
\cite{Simon,Zandi,Elston,Liebermeister,Metzler1,Metzler2,Orsogna,Abdolvahab1,Metzler3,Abdolvahab2,Krapivsky} specifically devoted to the chain translocation in the presence of chaperones.

In the pioneering study by Simon \textit{et al.} \cite{Simon}, the role of chaperones has been recognized as a
Brownian ratchet where the effect of chaperone's binding to the protein is to prohibit its backward diffusion through
the pore and consequently speed up the translocation. This mechanism rectifies the dynamics of the protein
compared with pure diffusion case.
Later, Zandi \textit{et al.} \cite{Zandi} investigated the chaperone-assisted translocation of a stiff polymer using
Brownian dynamics and proposed a new mechanism. At the moment when binding event occurs, the chain experiences a net
force along its length, pulling the chain. This result demonstrates that the role of chaperone is quite different from
that of a Brownian ratchet.
In a more detailed theoretical investigation based on master equation, Ambj\"{o}rnsson \textit{et al.} \cite{Metzler1,Metzler2}
examined chaperone-driven translocation of a stiff polymer. They identified three limiting dynamical regimes according
to binding situations: slow binding where either the chaperone concentration is low or binding strength is small (diffusive regime), fast binding but slow unbinding in which the chain cannot slide backwards (the irreversible binding regime), and fast binding and unbinding (the reversible binding regime).

However, the aforementioned studies did not take into account the chain flexibility due to the difficulties in theoretical treatment. Particularly, the chaperone's intersegment binding will lead to the bending of the polymer, thus the issue of chain flexibility is clearly important for chaperone-assisted translocation.
The basic questions associated with this process are the following: (a) What's the effect of the chaperone concentration and the binding energy on the translocation probability? (b) How do the chaperone concentration and the binding energy affect the translocation time? Is there an optimum chaperone concentration or an optimum binding energy for translocation?
These questions are very complicated and still not clear. To this end, using Langevin dynamics we investigate chaperone-assisted translocation of a flexible polymer through a nanopore.

\section{Model and methods} \label{chap-model}

\begin{figure}
\includegraphics*[width=\narrowfigurewidth]{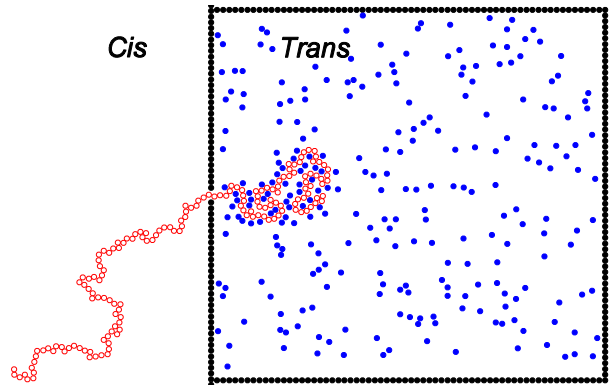}
\caption{(Color online) Schematic illustration of polymer translocation through a nanopore in the presence of chaperones
depicted by blue particles. The width of the pore is $w=1.6\sigma$.
        }
\label{Fig1}
\end{figure}

In our simulations, the polymer chain is modeled as a bead-spring chain of Lennard-Jones (LJ) particles with
the Finite Extension Nonlinear Elastic (FENE) potential. Excluded volume interaction between monomers is modeled
by a short range repulsive LJ potential: $U_{LJ}(r)=4\epsilon_0[{(\frac{\sigma}{r})}^{12}-
{(\frac{\sigma}{r})}^6] +\epsilon_0$ for $r\le2^{1/6}\sigma$ and 0 for $r>2^{1/6}\sigma$.
Here, $\sigma$ is the diameter of a bead, and $\epsilon_0$ is the depth of the potential.
The connectivity between neighboring monomers is modeled as a FENE spring with
$U_{FENE}(r)=-\frac{1}{2}kR_0^2\ln(1-r^2/R_0^2)$, where $r$ is the distance between consecutive monomers,
$k$ is the spring constant and $R_0$ is the maximum allowed separation between connected monomers.

We consider a two-dimensional geometry as depicted by \ref{Fig1}, where the chaperones of diameter $\sigma$ are modeled as mobile beads moving within the rectangular box with a pore of width $w=1.6\sigma$ formed by stationary wall particles of the same size.
The wall particle interacts with the monomer and the chaperone by the repulsive LJ potential as shown above, while
chaperone-monomer interaction is modeled by an attractive LJ potential with a cutoff of 2.5$\sigma$ and interaction
energy $\epsilon$.
In the Langevin dynamics simulations, each monomer is subjected to conservative, frictional and random forces, respectively.
Namely, $m{\bf \ddot{r}}_i=-{\bf\nabla}({U}_{LJ}+{U}_{FENE})-\xi {\bf v}_i+{\bf F}_i^R$, where $m$ is
the bead's mass, $\xi$ is the friction coefficient, ${\bf v}_i$ is the bead's velocity, and ${\bf  F}_i^R$ is the random force which satisfies the fluctuation-dissipation theorem \cite{Allen}.

In our model, the LJ parameters $\epsilon_0$, $\sigma$ and bead mass $m$ fix the system energy, length and mass scales, leading to the corresponding time scale $t_{LJ}=(m\sigma^2/\epsilon_0)^{1/2}$.
Each bead corresponds to a Kuhn length of a polymer, so  we choose $\sigma \sim 1.5$ nm and the bead mass $m \approx 936$ amu \cite{Luo1}.
We set $k_{B}T=1.2\epsilon_0$, which means the interaction energy $\epsilon_0$ to be $3.39 \times 10^{-21}$ J at actual temperature 295 K. This leads to a time scale of 32.1 ps \cite{Luo1}.
The dimensionless parameters in our simulations are chosen to be $R_0=1.5$, $k=30$, $\xi=0.7$.
Then the Langevin equation is integrated in time by the method proposed by Ermak and Buckholz \cite{Ermak}.
Initially, the first monomer is placed at the $trans$ side with one unit length to the pore center.
Then, the remaining monomers and chaperones are under thermal collisions described by the Langevin thermostat to
reach equilibrium state of the system.
Typically, each simulation data is the average result of 1000 successful translocation events to minimize statistical
errors.

The current simulation method was also used to investigate polymer translocation driven in the presence of a cross-membrane electrical potential \cite{Luo,Luo4}.

\section{Results and discussions} \label{chap-results}

Based on master equation,  Ambj\"{o}rnsson \textit{et al.} \cite{Metzler1,Metzler2} investigated the dynamics of chaperone-driven translocation of a stiff polymer. Although polymeric degrees of freedom of the translocating chain is neglected and the system is assumed to be close to equilibrium, their results still can shed light on the translocation of the flexible polymer.

For the chaperone binding to the binding site, the probability that a binding site is occupied depends on the chaperone concentration $c_0$ and the binding energy $\epsilon$. By calculating the binding partition function, they obtain a dimensionless parameter $\kappa=c_0K^{eq}$ as a relevant measure of the effective binding strength \cite{Metzler1,Metzler2}, where the equilibrium binding constant $K^{eq}=v_0 exp({\epsilon/k_BT})$ with  $v_0$ being the typical chaperone volume. For univalent binding, the equilibrium probability that a binding site is occupied is
\begin{equation}
P_{occ}^{eq}=\kappa/(1+\kappa).
\label{eq1}
\end{equation}

For stiff polymers, the force acting on polymer, $F_{bind}(s)$, in unit of $k_BT/\sigma$ for reversible binding of chaperones to the translocating polymer is \cite{Metzler1}
\begin{equation}
F_{bind}(s)=ln(1+\kappa).
\label{eq2}
\end{equation}
where $s$ is the translocation coordinates. This results indicates that the force increases with the chaperone concentration and the binding energy and is independent of $s$.

For flexible polymers, we don't expect that the effective binding strength $\kappa$ and the probability that a binding site is occupied $P_{occ}^{eq}$ are the same as those for stiff polymers. Due to the chain flexibility, one chaperone can attach to several binding sites (intersegmental binding) for the high chaperone concentration and binding energy as observed in the simulation, leading to chain folding. But $\kappa$ and $P_{occ}^{eq}$ should also increase with the chaperone concentration and the binding energy.

In addition, for flexible polymers entropic effects from \textit{cis} and \textit{trans} sides give an additional contribution to the total force. As noted by Ambj\"{o}rnsson \textit{et al.} \cite{Metzler1}, the binding force and the entropic force are additive if the chaperone binding is independent of the curvature of the polymer.
For flexible polymers, we can write the total force acting on the translocating polymer as
\begin{equation}
F(s)=F_{bind}(s)- F_{trans, e}(s,c_0,\epsilon)-F_{cis, e}(s).
\label{eq3}
\end{equation}
Here, $F_{bind}(s)$ is the force from the binding which should increase with the chaperone concentration and the binding energy, although its explicit expression may be different from \ref{eq2} for high binding energy.
$F_{trans, e}(s,c_0,\epsilon)$ is the entropic force from the \textit{trans} side, which also increases with the chaperone concentration and the binding energy, resulting from either crowding effect induced by high chaperone concentration or the intersegmental binding. The entropic force from the \textit{cis} side $F_{cis, e}(s)$ plays an important role only at the beginning of the translocation, and it is negligible compared with $F_{trans, e}(s,c_0,\epsilon)$ particularly for high chaperone concentration and binding energy. Therefore, the translocation dynamics is dominated by the interplay between $F_{bind}(s)$ and $F_{trans, e}(s,c_0,\epsilon)$.

In our simulation, the chaperone concentration is proportional to the ratio of the number of the chaperone ($N_c$) to the constant area of the rectangular box ($64\times64$). In the following, we use $N_c$ to stand for the chaperone concentration.

\subsection{The effect of $\epsilon$ and $N_c$ on the translocation probability}

\begin{figure}
\includegraphics*[width=\figurewidth]{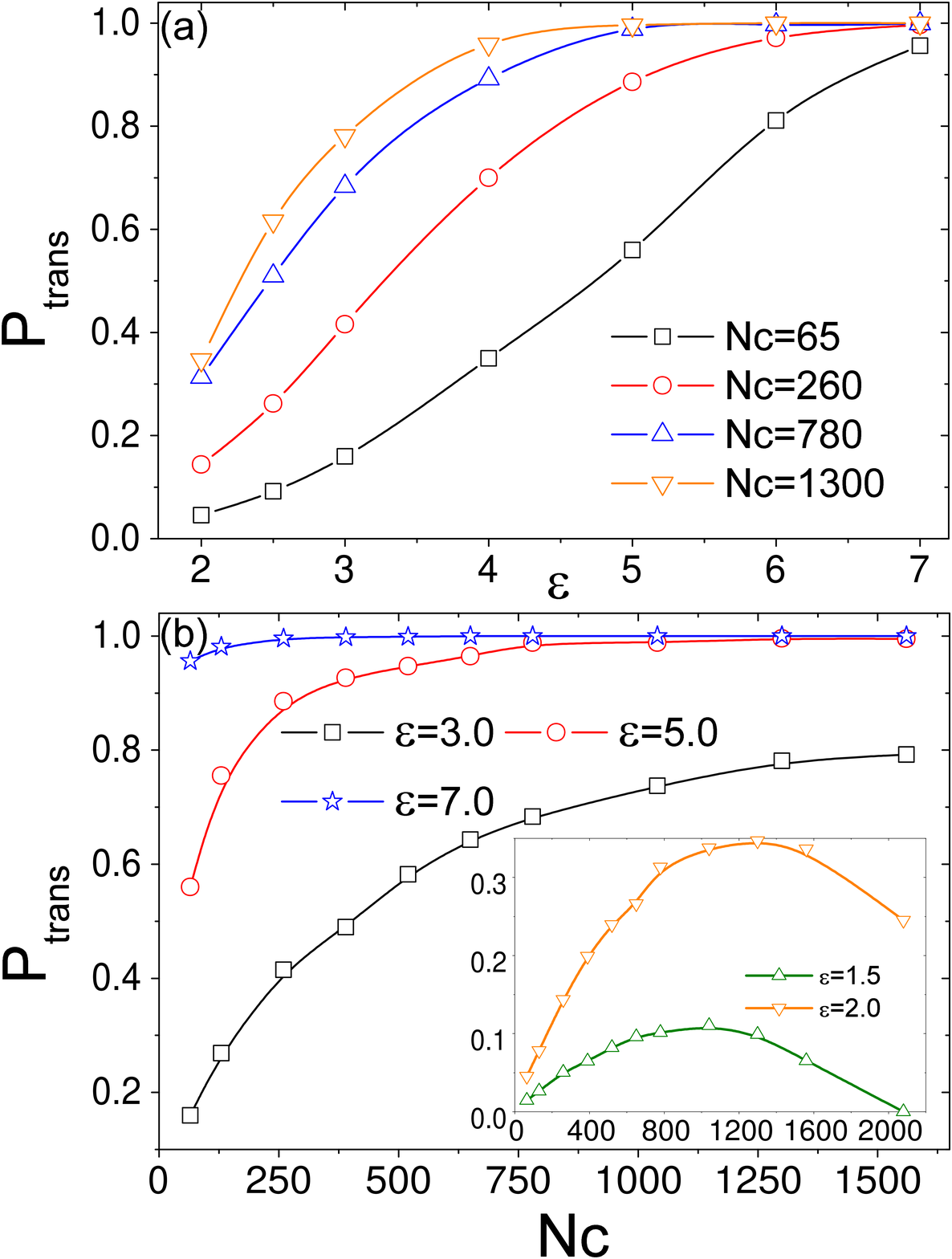}
\caption{The translocation probability as a function of (a) the binding energy $\varepsilon$ for different
chaperone concentration ($\sim N_c$) and (b) the chaperone concentration for different binding energy. The chain
length $N=64$.
The inset shows situations for weak binding.
        }
\label{Fig2}
\end{figure}

The translocation probability, $P_{trans}$, refers to the ratio of the successful translocation events to the whole attempts at given physical parameters in the simulation.
With increasing $\epsilon$, $P_{trans}$ increases rapidly first, and then slowly approaches saturation at larger
$\epsilon$ for higher chaperone concentration, while it continuously increases for lower $N_c$,
as shown in \ref{Fig2}(a).

The observed $P_{trans}$ can be well understood by taking into account the force in \ref{eq3} as a function of $N_c$ and $\epsilon$. The force exerted on the chain by chaperone's binding, $F_{bind}(s)$, increases with increasing the chaperone concentration and the binding energy $\epsilon$. For high binding energy $\epsilon$, $F_{bind}(s)$ is the dominant term in \ref{eq3} for the translocation process. Meanwhile, unbinding is slower, as a consequence the chaperone acts as a ratchet which would effectively prohibit the chain's backward motion out of the pore. Qualitative, $P_{trans}$ shows the similar behaviors as the probability that a binding sites is occupied, see $P_{occ}^{eq}$ in \ref{eq1}.

\ref{Fig2}(b) shows that $P_{trans}$ also goes up rapidly first with increasing $N_c$ and then slowly approaches the
saturation at higher $N_c$ for strong binding. The reason for the increase of $P_{trans}$ with $N_c$ is that the probability of chaperone's binding to the chain rises with increasing $N_c$.
However, for weak binding $P_{trans}$ shows a maximum with increasing $N_c$, as shown in the inset of \ref{Fig2}(b).
This unexpected decrease for high chaperone concentration stems from crowding effect at the \textit{trans} side. Obviously, $F_{bind}(s)$ increases with $N_c$, however, $F_{trans, e}(s,c_0,\epsilon)$ also increases with growing $N_c$ due to fact that the chain has to overcome greater entropic barrier for translocation and becomes dominant. Moreover, for higher chaperone concentration, the collision frequency between chaperones increases, which leads to higher unbinding events particularly for weak binding. The interplay of these factors results in the maximum of $P_{trans}$.

\subsection{Distribution of the translocation time}

\begin{figure}
\includegraphics*[width=\figurewidth]{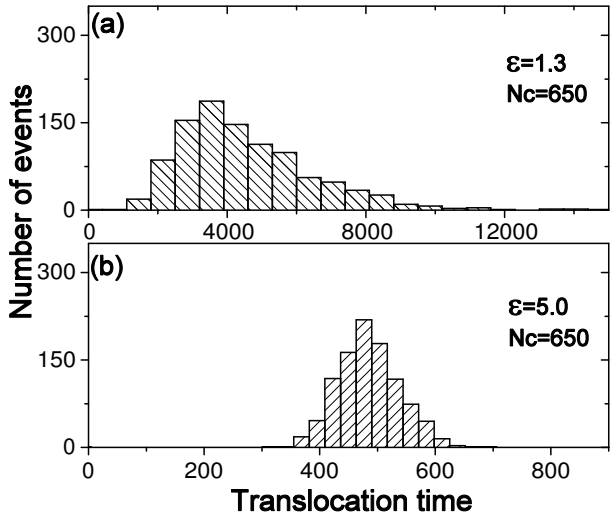}
\caption{(Color online) The distribution of translocation time for different binding energy under $N_c=650$. The chain
length is $N=64$.
        }
\label{Fig3}
\end{figure}

We also checked that the binding energy $\epsilon$ has an obvious effect on the shape of the histogram of the translocation time, as shown in \ref{Fig3}.
For $N=64$ and $N_c=650$, the distribution of translocation time is asymmetric with a long exponential tail for weak binding $\epsilon=1.3$, while for strong binding $\epsilon=5.0$, it nearly approaches a Gaussian distribution.
The reason is that with increasing $\epsilon$ the force from binding $F_{bind}(s)$ greatly increases and is the dominant term in the total force $F(s)$.
For driven translocation by cross-membrane electric field, we also observed similar distributions for very weak and strong driving forces, respectively \cite{Luo2}.

\subsection{Translocation time as a function of $\epsilon$ and $N_c$}

\begin{figure}
\includegraphics*[width=\figurewidth]{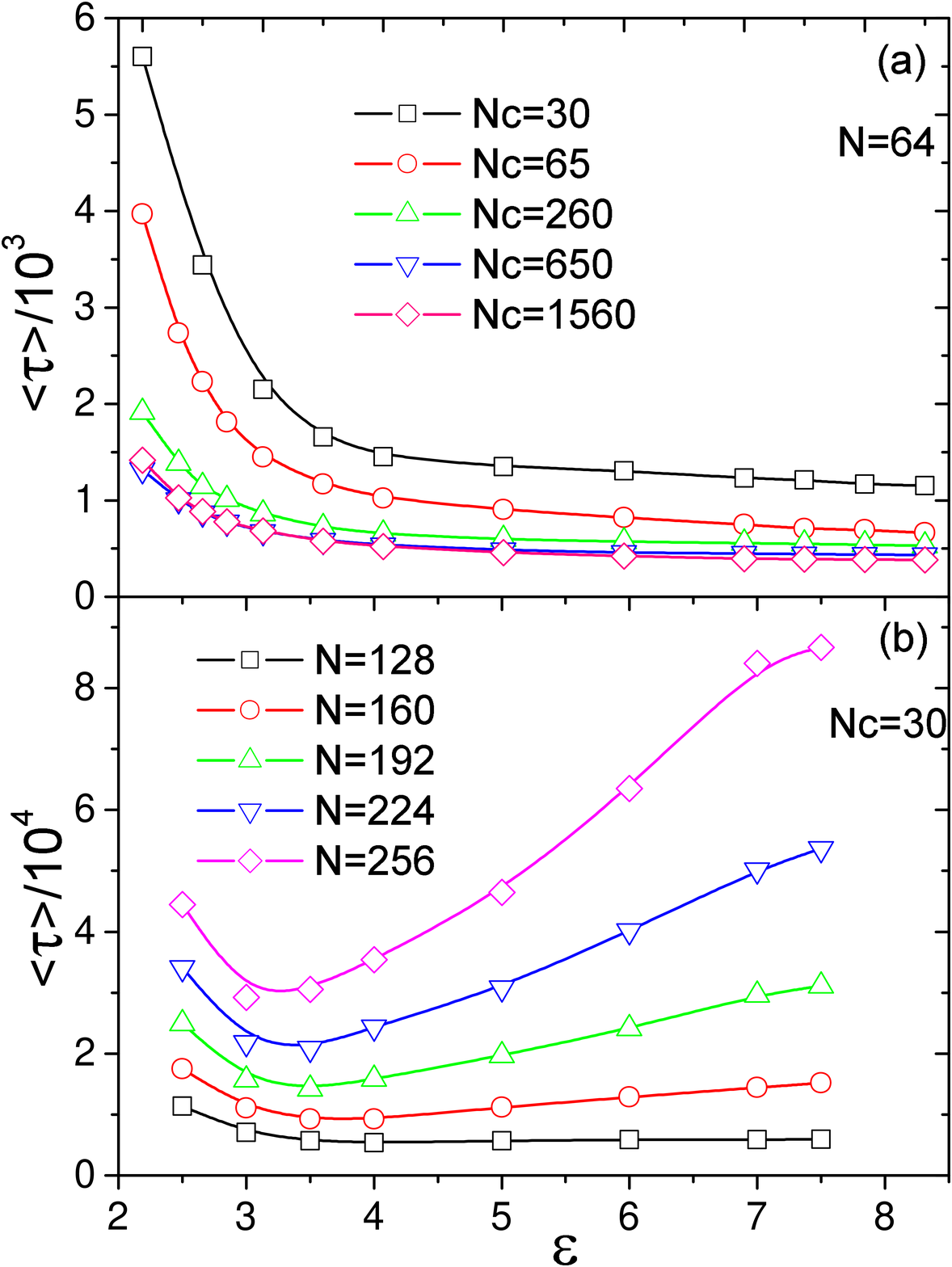}
\caption{(Color online) Translocation time as a function of the binding energy for (a) $N=64$ and different chaperone
concentration, and (b) $N_c=30$ and different chain length $N$.
        }
\label{Fig4}
\end{figure}

\ref{Fig4}(a) shows $\tau$ as a function of $\epsilon$ for $N=64$ and different $N_c$. We find that $\tau$ decreases with increasing $\epsilon$ (at least for $\epsilon\leq8.5$). It is not difficult to understand the overall fall of $\tau$ with $\epsilon$ from the perspective of the magnitude of $F(s)$ and the ratchet mechanism that prevents back-sliding motion as noted above.  Increasing $\epsilon$ leads to the increase of $F_{bind}(s)$, and is beneficial to the forward motion of the chain, speeding up the translocation.
What draws our attention most is the notable discrepancy in the decay rate of $\tau$, i.e., $\tau$ decreases quite rapidly with $\epsilon$ for $\epsilon$ lower than a critical binding energy $\epsilon_c$, but much slowlier at $\epsilon>\epsilon_c$ regimes.
%
Specifically, for $\epsilon<\epsilon_c$, the binding frequency is higher but the force $F_{bind}(s)$ from the binding is smaller compared with the case of strong binding. The latter dominates the translocation dynamics, leading to the observed behavior. For $\epsilon>\epsilon_c$, translocation is slowed due to the competition between $F_{bind}(s)$ and the entropic force $F_{trans, e}(s,c_0,\epsilon)$, leading to slower increase of $F(s)$ with increasing $\epsilon$. On the one hand, $F_{bind}(s)$ increases with $\epsilon$. On the other hand, intersegmental binding between translocated monomers results in larger entropic force $F_{trans, e}(s,c_0,\epsilon)$ due to the loss of chain conformations.

However, for longer chain length $N\ge128$ and lower chaperone concentration $N_c=30$ as shown in \ref{Fig4}(b), we observed an optimum $\epsilon$ for translocation, namely $\tau$ has a minimum as a function of $\epsilon$.
The time $\tau_{diff}$ for a polymer of length $N$ to diffuse a distance of the order of the binding site length $\sigma$ is $\tau_{diff}\sim \sigma^2/D \sim N\xi\sigma^2/(k_BT)$ with the diffusion constant of the chain $D=k_BT/(N\xi)$.
The average distance between chaperones in solution is $R_c = L/\sqrt{N_c}$, where $L=64$ is the length of the simulation box. It is suffice for a chaperone to diffuse a distance of the order of $R_c$ for any one chaperone to attach the binding sites (provided the binding energy is sufficiently high), and this time is $\tau_{unocc}= R_c^2/4D_c= \xi \sigma^2 L^2/(k_BTN_c)=34.1\xi \sigma^2/(k_BT)$, where the diffusion constant of chaperones $D_c=k_BT/\xi$ due to the same size for a chaperone and a monomer.
Due to diffusion through a nanopore, $\tau_{diff}$ has a large prefactor \cite{Chuang}. Thus, $\tau_{diff}$ is much longer than $\tau_{unocc}$ for larger $N$, which indicates that it is possible for chaperones to bind the chain very soon by taking into account the range of interaction of the cutoff $2.5\sigma$ used in the simulation. For chain length $N < 128$, the chaperone can almost cover all the binding sites for large $\epsilon$ by the intersegmental binding, thus there is no minimum for translocation time. But for longer chains, all chaperones have become bound and there are no free chaperones left before completing the translocation, leading to the increase of the translocation time.

\begin{figure}
\includegraphics*[width=\figurewidth]{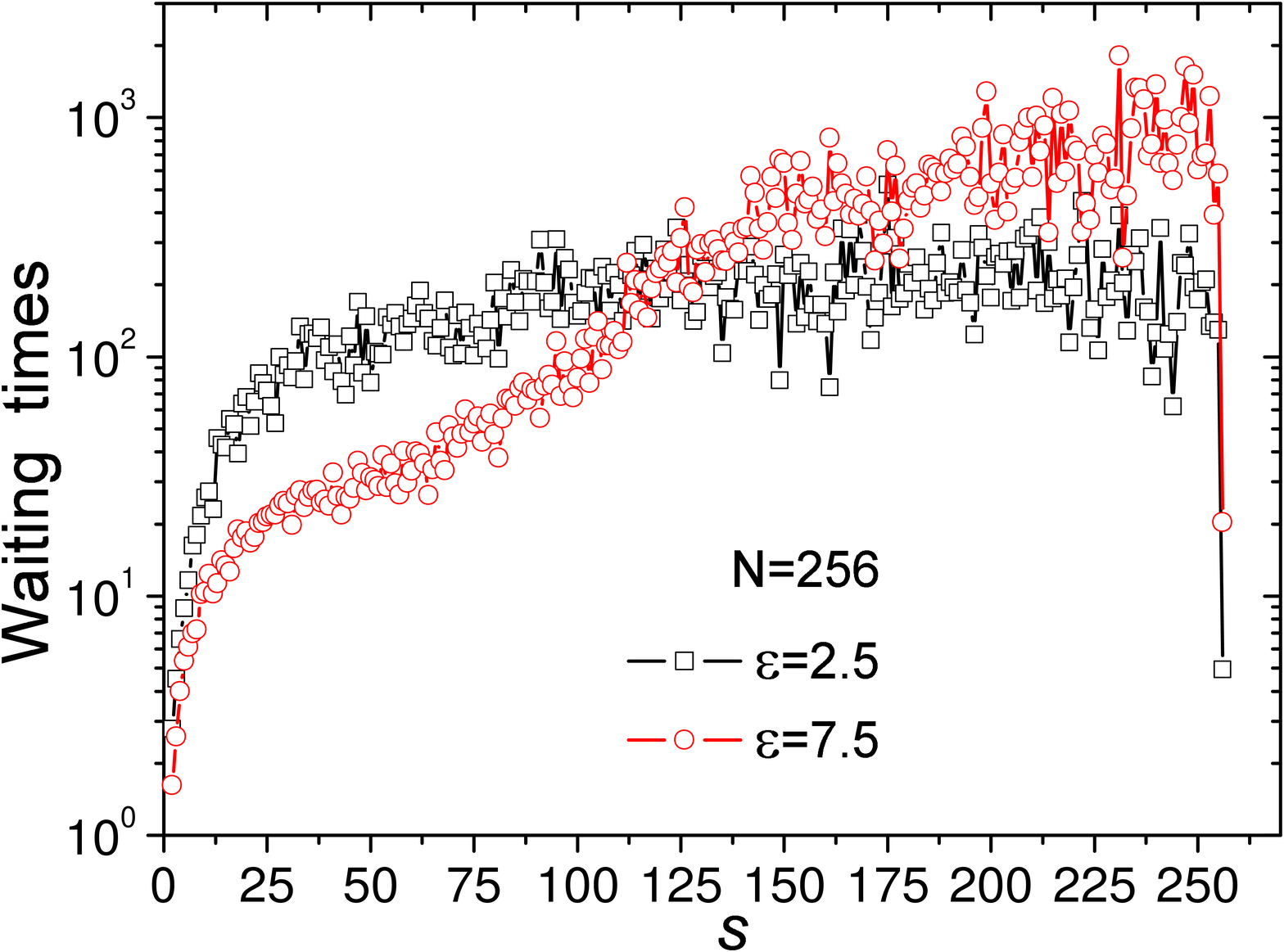}
\caption{Waiting time distribution for $N=256$, $N_c=30$ and different binding strength $\epsilon$.
        }
\label{Fig5}
\end{figure}

To understand this unexpected behavior, we have also investigated the distribution for waiting (residence) time of bead $s$, which is defined as the time between the events that the bead $s$ and the bead $s+1$ exit the pore. \ref{Fig5} shows the waiting time distribution for $N=256$, $N_c=30$ and different binding strength $\epsilon=2.5$ and 7.5, respectively.
For $\epsilon=2.5$, the waiting time increases rapidly first and then approaches the saturation very soon, indicating
the feature of fast binding and unbinding. However, for $\epsilon=7.5$ the waiting time always increases except for last several monomers.
What's more, the translocation is faster for $\epsilon=7.5$ than for $\epsilon=2.5$ when $s\lesssim125$, however, it is much slower for $\epsilon=7.5$ than for $\epsilon=2.5$ when $s\gtrsim125$. All together, the translocation time for $\epsilon=7.5$ is much longer than that for $\epsilon=2.5$. The reason for observed behavior for $s\gtrsim125$ is that, for $\epsilon=7.5>\epsilon_c$ and low chaperone concentration, almost all chaperones keep bound to the front part of the chain, rendering that there is no free chaperones for latter new-emerging segments and the corresponding translocation becomes diffusive \cite{Chuang}.

\begin{figure}
\includegraphics*[width=\figurewidth]{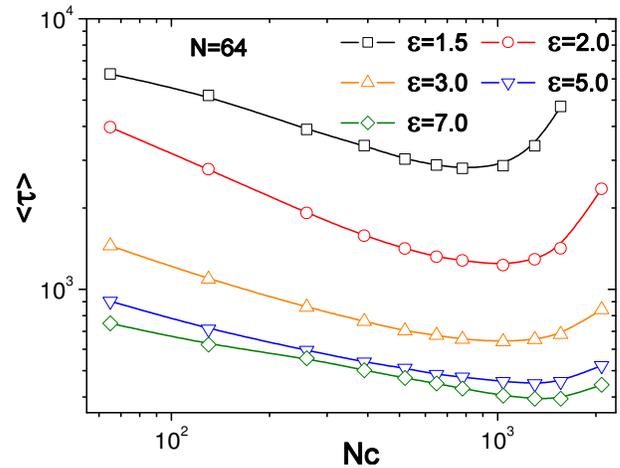}
\caption{(Color online) The influence of chaperone concentration on the translocation time for different $\epsilon$. The chain length is $N=64$.
        }
\label{Fig6}
\end{figure}

\ref{Fig6} shows $\tau$ as a function of $N_c$ for different $\epsilon$. We find that $\tau$ initially decreases and
subsequently goes up with increasing $N_c$. As stated above, growing $N_c$ leads to an increase in $F_{bind}(s)$ as well as $F_{trans, e}(s,c_0,\epsilon)$. Considering that $F_{bind}(s)$ is the dominant term in \ref{eq3} for initial increasing of $N_c$, the total force $F(s)$ increases, resulting in the decrease of $\tau$.
Besides, the probability that the binding site just passing through the pore exit also increases with growing $N_c$,  which is favorable to the formation of the ratchet.

However, overfull chaperones gives rise to the crowding effect and the entropic force $F_{trans, e}(s,c_0,\epsilon)$ prevails in the competition with $F_{bind}(s)$, hindering the translocation and resulting in the increase in $\tau$ instead of continuous fall. Moreover, the corresponding chaperone concentration of the minimum shifts to higher $N_c$ value with increasing $\epsilon$. This is due to the fact that with increasing $\epsilon$, the superiority of $F_{trans, e}(s,c_0,\epsilon)$ works for higher $N_c$ and the formed ratchet could prevent the back-sliding motion of the chain more effectively. Needless to say, it is conducive for the translocation of the chain through the nanopore and thus postpones the advent of the minimum point.

\subsection{Translocation time as a function of the chain length}

\begin{figure}
\includegraphics*[width=\figurewidth]{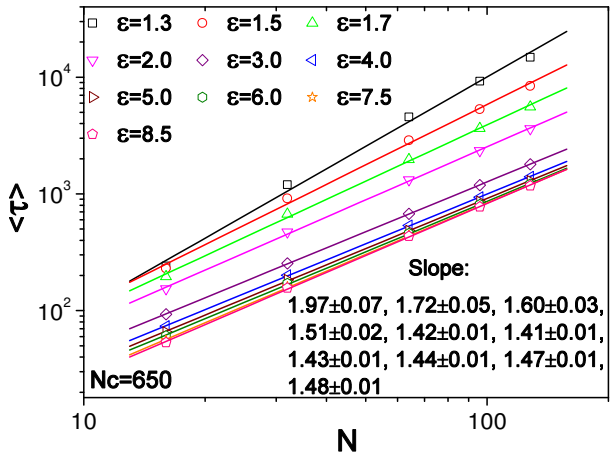}
\caption{(Color online)  Translocation time as a function of the chain length $N$ for different binding energy $\epsilon$ under fixed chaperone concentration $N_c=650$.
        }
\label{Fig7}
\end{figure}

Similar with the electric-field driven case, the presence of chaperones only in the $trans$ side could also induce a chemical potential difference between the two sides of the membrane. The scaling of the translocation time $\tau$ with the chain length $N$ is an important measure of the underlying dynamics, $\tau \sim N^\alpha$ with $\alpha$ being the scaling exponent. Our previous two dimensional electric-field driven translocation simulations \cite{Luo2,Luo3} show there was a crossover from $\alpha=2\nu_{2D}$ for fast translocation to $\alpha=1+\nu_{2D}$ for slow translocation, where $\nu_{2D}=0.75$ is the Flory exponent for a self-avoiding chain in two dimensions \cite{deGennes}. Most recently, we further find that for fast translocation processes $\alpha=1.37$ in three dimensions, while it crosses over to $\alpha=1+\nu_{3D}$ with $\nu_{3D}=0.588$ for slow translocation, corresponding to weak driving forces and/or high friction \cite{Luo4}.

\ref{Fig7} shows the $\tau$ as a function of  $N$ for moderate chaperone concentration $N_c=650$ and different $\epsilon$.
Obviously, the scaling exponent $\alpha$ depends significantly on $\epsilon$: it initially decreases from $1.97\pm0.07$ to a minimum $1.41\pm0.01$, following by a slight increase with increasing $\epsilon$.
For higher $\epsilon$, the force from the binding $F_{bind}(s)$ is greater and dominates the translocation dynamics, leading to $\alpha \approx 2\nu_{2D}$ for fast translocation process as the electric-field driven translocation \cite{Luo2,Luo3}. Decreasing $\epsilon$ to $\epsilon=1.3$, $\alpha$ increases to 1.97. If decreasing $\epsilon$ further, it may access unbiased translocation regime with $\alpha=N^{1+2\nu_{2D}}$ \cite{Chuang}.

\section{Conclusions} \label{chap-conclusions}

Using Langevin dynamics simulations, we investigate the dynamics of chaperone-assisted translocation of a flexible
polymer through a nanopore. We find that increasing the binding energy $\epsilon$ between the chaperone and the chain and the chaperone concentration can greatly improve the translocation probability. Particularly, with increasing
the chaperone concentration a maximum translocation probability is observed for weak binding. For a fixed
chaperone concentration, the histogram of translocation time $\tau$ has a transition from long-tailed
distribution to Gaussian distribution with increasing $\epsilon$. $\tau$ rapidly decreases and then almost saturates
with increasing binding energy for short chain, however it has a minimum for longer chains at lower chaperone
concentration. We also show that $\tau$ has a minimum as a function of the chaperone concentration. For different $\epsilon$, a nonuniversal dependence of $\tau$ on the chain length $N$ is also observed. These results can be interpreted by characteristic entropic effects for flexible polymers induced by either the crowding effect from high chaperone concentration or the intersegmental binding for the high binding energy.

Generally, chaperones have a size larger than that of a monomer, giving rise to the ``parking lot effect'' as observed by previous studies \cite{Metzler1,Metzler2,Orsogna,Abdolvahab1,Metzler3,Abdolvahab2}, leading to a less efficient translocation: after binding of a chaperone to the chain close to the pore exit, the chain firstly needs to diffuse by a chaperone size distance, before next binding event occurring. In addition, we have also assumed that the binding energy is the same along the chain. However, proteins and nucleic acids consist of heterogeneous sequence of aminoacids, bases or base pairs, respectively. It has been found that the chain heterogeneity is important in translocation dynamics for stiff polymers \cite{Abdolvahab2}. In the future studies, it would be interesting to investigate the effects of the size difference, the changes in chain flexibility and the chain heterogeneity along the chain on the translocation dynamics.


\begin{acknowledgement}
This work is supported by the ``Hundred Talents Program'' of Chinese Academy of Science (CAS) and the National Natural Science Foundation of China (Grant No. 21074126).
\end{acknowledgement}
\begin{suppinfo}
Complete ref 4.
\end{suppinfo}

\begin{tocentry}

\includegraphics*[width=6.0cm,height=3.5cm]{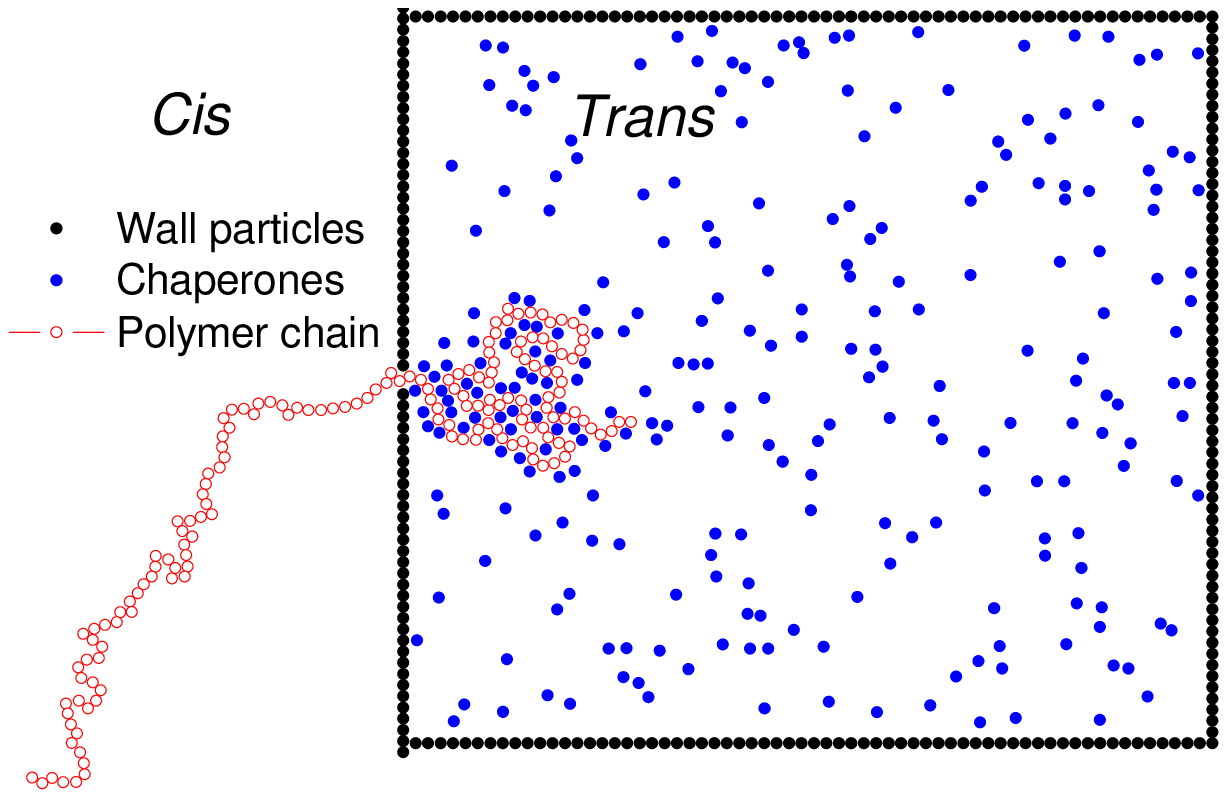}

\end{tocentry}

\end{document}